\documentclass[aps,prb,floatfix,
  prb,
 groupedaddress]{revtex4}

\usepackage{graphicx}%
\usepackage{dcolumn}
\usepackage{amssymb,amsmath}
\usepackage{bm}
\usepackage[usenames]{color}

\begin{document}

\author{Oleg L. Berman, Roman Ya. Kezerashvili, and German V. Kolmakov}

\affiliation{Physics Department, New York City College of Technology,\\ City University of New York, Brooklyn, NY 11201, USA}

\title{Polariton-based  optical switch}

\begin{abstract}
Based on the studies of  propagation of an exciton-polariton condensate
in a patterned optical microcavity with an embedded graphene layer, we propose
a design of a Y-shaped electrically-controlled optical switch. The polaritons
are driven by a time-independent force due to the microcavity wedge shape and by
a time-dependent drag force owing to the  interaction of excitons in graphene
and the electric current running in a neighboring quantum well. It is demonstrated that
one can control the polariton flow direction by changing  the direction
of the electric current. The simulations also show that an
external electric field normal to the microcavity plane can be utilized as an
additional parameter that controls the propagation of the signals in the switch.
By considering the transient dynamics of the polariton condensate,
we estimate the response speed of the switch.
Finally, we propose a design of the polariton switch in a flat microcavity
based on the geometrically  identical Y-shaped  quantum well and graphene pattern where the
polariton flow is only induced by the drag force.
\end{abstract}

\maketitle

\section{Introduction}\label{sec:intro}

In the past decades, substantial experimental and theoretical
efforts  were devoted to find the optimal ways to tune the optical
properties of semiconductors and graphene by application of external
electric and magnetic fields. The motivation of this research lies
in the potential applications for integrated circuits in optical and
quantum computers, for secure information transfer, and in new light
sources.\cite{Gibbs:11} One of the promising approaches is in the
use of polaritons, which are a quantum superposition of cavity
photons and excitons in a nanometer-wide semiconductor layer (a
quantum well) or
graphene.\cite{Zamfirescu:02,Deng:03,Liew:10,Menon:10,Deng:10,Berman:12c}
Since polaritons are interacting Bose particles, a polariton gas can
transit to a superfluid state that, under certain conditions,
propagates in a microcavity almost without
dissipation.\cite{Carusotto:04,Amo:09,Wouters:10a,Berman:12c} Due to
a small effective mass,  $10^{-4}$ of the free electron mass, the
superfluid transition occurs at relatively high temperatures that
are comparable with the room temperature. Polaritons can propagate
in the sample with a  speed up to a few percent of the speed of
light that is, much faster than that of electric-field driven
electrons and holes in semiconductors.\cite{Wouters:10b}  However,
one of the main problems to be solved in actual design of
polariton-based optical devices is in weak response of polaritons to
an external electric field owing to their net zero electric charge.

\begin{figure*}[t]
\includegraphics[width=7cm]{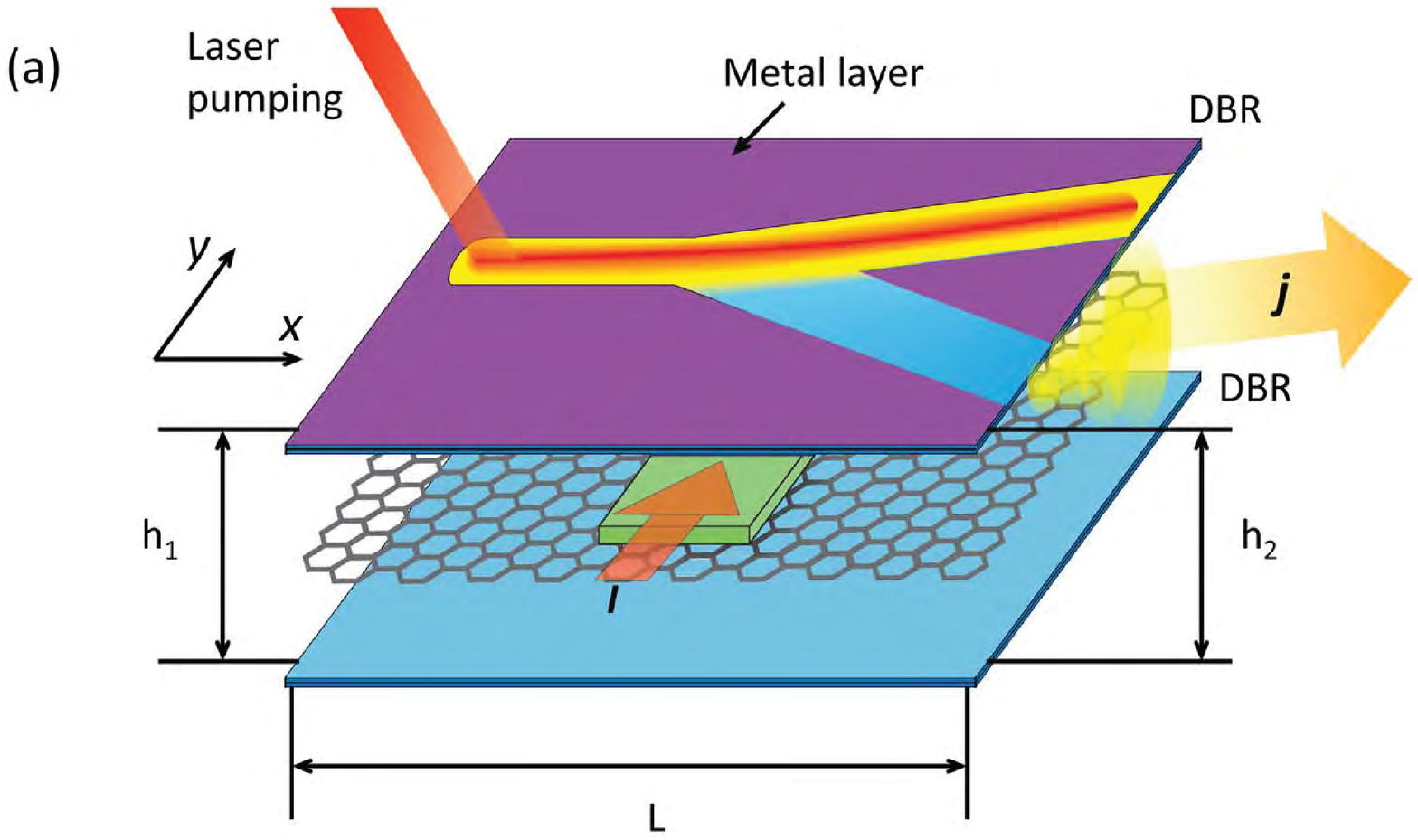}\vspace{0.5cm}
\includegraphics[width=7cm]{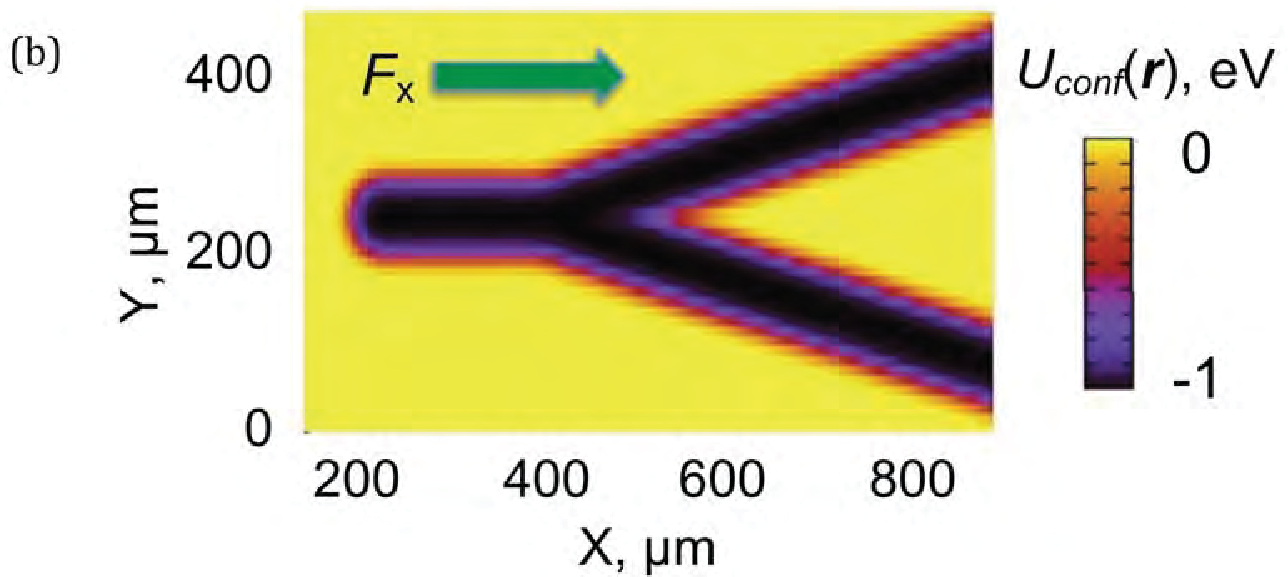}
\caption{(a) Schematic of the wedge-shaped microcavity formed by two
distributed Bragg reflectors (DBR) that encompasses an embedded
graphene layer. Polaritons in the microcavity are formed as a
quantum superposition of excitons in graphene and cavity photons.
The force $F_x$  is exerted on the polaritons in $x$-direction owing
to a small opening angle $\alpha \approx (h_2-h_1)/L \ll 1$ of the
microcavity, where $h_1$ and $h_2$ are the microcavity lengths at
the opposite edges and $L$ is the horizontal size of the
microcavity. A metal layer deposited on the upper DBR creates a
Y-shaped potential energy landscape for the polaritons shown in (b).
The polaritons are created by an external laser radiation in the
stem of the channel and propagate towards the junction due to the
force $F_x$. A driving electric current $I$ runs perpendicularly to
the stem of the channel in a quantum well QW (green) in the area of
the junction. The polariton flux $\bm{j}$ is formed in one branch of
the channel in response to the drag created by the current. }
\label{fig1}
\end{figure*}

\begin{figure*}[t]
\includegraphics[width=7cm]{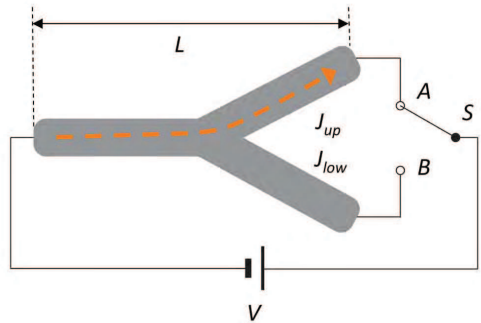}
 \caption{Schematic of the polariton switch where the polariton
flux in a flat microcavity is induced by the drag  due to an
electric current
running in the  circuit.  
 Both the quantum well that carries the electron current and the graphene layer containing the excitons
 have identical geometrical Y shape and are positioned parallel to each other.
 The diffraction Bragg reflectors positioned parallel to the graphene layer and the quantum well are not shown.
   The position of switch $S$ controls the direction of the polariton flow.}
\label{fig6}
\end{figure*}

In this paper,  we propose a design of polariton-based, electrically
controlled  switch, in which the polariton propagation is controlled
by means of a  drag force.  Recently, it was shown that the drag
caused by an electric current results in entrainment of polaritons
and in formation of a persistent, directed polariton current in the
microcavity.\cite{Berman:10a,Berman:10b} In this case, this is a
drag force that is exerted  on the exciton component of polaritons
that results in  generation of the polariton flux proportional to
the driving electric field $\bm{E}$,
\begin{equation}
\bm{j} = \gamma \bm{E}, \label{eq:j0}
\end{equation}
where  $\gamma$ is the temperature-dependent drag coefficient.
In addition to the electrically controlled drag force, the polaritons can also be accelerated by a constant,
time independent force, which is
caused by  the wedge-like shape of the microcavity.\cite{Sermage:01,Nelsen:13}
In the latter case, the effect of the constant force is similar to application of a dc voltage in
electric circuits. By changing the magnitude and direction of the electric current
one can vary the net force exerted on the polaritons and thus, to control the polariton propagation.

In this paper, we propose  the polariton switches of two types. In
the first design, the polariton flow in the switch is produced by a
constant force due to the wedge-like shape of the microcavity
whereas the drag force is utilized to switch the direction of the
flow. In the second design, the polariton flow in the required
direction is only produced by the drag force. The photons propagate
in the planar microcavity, and the excitons are located in a quantum
well embedded in the microcavity. Below we describe the both types
of the switches in details.

In the first design, we take advantage of a patterned microcavity to
create a potential-energy landscape for polaritons in the form of
one-dimensional channels, or ``polariton wires'', which are somewhat
similar to the ordinary wires that conduct electrical current.
However, in our case, the wires  conduct polaritons and help one to
deliver photons to a desired location in the semiconductor
structure. Recently developed experimental techniques, which enable
one to produce such guiding potentials for polaritons, include
deposition of a metal pattern on the Bragg
reflectors\cite{Utsunomiya:08} and modulation of the cavity layer
thickness in a mesa structure.\cite{Daif:06,Kaitouni:06} The
polariton channels of complex shape and topology can also be
engineered directly in an air-filled microcavity by creating
quasi-one-dimensional waveguides, or
micropillars.\cite{Wertz:10,Vugt:11,Nguyen:13,Das:13,Boulier:14}

To design a polariton switch, we utilize a Y-shaped channel  shown
in Fig.~\ref{fig1}. We assume that polaritons are  created at a
constant rate by the external laser pumping at the stem of the
channel and then, propagate towards the junction under the action of
a constant force due to the microcavity wedge-like shape. After
reaching the junction, the polariton flow splits between the two
branches. If no drag force is applied to the polaritons, the
 flux  is distributed equally between the branches. However, if the drag force is exerted on the polaritons
in the region of the junction in the direction normal to the stem,
the polaritons are ``pushed'' towards one branch and hence, the flux
in that branch increases. Simultaneously, the flux through another
branch decreases due to approximate conservation of the total number
of polaritons and in effect, the polariton flux is redistributed
between the branches. The  drag force in the junction region  is
produced by the electric current running in a neighboring quantum
well, which has a form of a stripe placed in a microcavity
perpendicular to the stem of the  channel. Thus, the heterostructure
encompasses two parallel quantum wells, separated by a
nanometer-wide semiconductor of dielectric barrier, placed in an
optical microcavity. Heterostructures embedded in a microcavity are
widely used in the studies of  exciton
dynamics\cite{Balili:09,Butov:03} and in experiments with
polaritons.\cite{Lai:07,Assmann:11,Das:13} We demonstrate below that
more than 70\% of the total polariton flux can be dynamically
redistributed between the branches for realistically achievable drag
forces.

The second type of the polariton switch is based on a flat microcavity where
the mirrors are parallel to each other.
In this design, the excitons are located in a Y-shaped semiconductor
quantum well,  and the neighboring current-carrying quantum well
geometrically repeats the shape of the structure, see
Fig.~\ref{fig6}. In contrast  to the first type of the switch
described above, here we do not use the patterned microcavity.
Instead, the polaritons are guided by the Y-shaped semiconductor
structure. The electric current in the quantum well is generated by
the external voltage.
The drag force exerted on the cavity polaritons by the electric current
is directed parallel to the channel and induces the polariton flow along one of the branches.
The polariton flow  is redistributed between the branches in
response to switching of the electric voltage applied to the current-carrying quantum well.

In both designs, the quantum wells containing the excitons can be
fabricated by using a number of materials, including
GaAs,\cite{Balili:09,Butov:03} CdTe,\cite{Kasprzak:06}
WSe$_2$,\cite{Wu:14} and gapped
graphene.\cite{Furchi:12,Youngblood:14}  While the polariton
condensate was experimentally observed when the polaritons are
formed by excitons in a semiconductor quantum well~\cite{Deng:10},
in this paper we focus on microcavity polaritons formed by excitons
in  graphene. One of the advantages to use graphene is related to
the properties of polaritons in gapped graphene that can be tuned by
application of a static electric field normal to the microcavity
plane. Specifically, the energy gap between the valence and
conduction bands can be created and changed by the electric field in
a wide range.\cite{Kuzmenko:09, Mak:09, Zhang:09} This enables one
to tune the polariton effective mass  and the sound velocity in the
spectrum of  the quasiparticles~\cite{Berman:10c} and through that
to optimize the performance of the system, as detailed below.
Another advantage of considering excitons in a graphene layer
compared to a semiconductor quantum well is the absence of the
fluctuations of the width due to the fact that graphene is a 2D one
atom-thick material. Such fluctuations cause the disorder which
results to the decrease of the drag coefficients due to the
scattering of the excitons on the random field.  Thus, in both
designs of the polariton switch described above we consider graphene
embedded in microcavity. Our consideration of the switch with
polaritons formation in a quantum well currently is in progress.

\section{The dynamics of a polariton condensate}

\label{sgpe}

The dynamics of the polariton condensate was captured via
 the non-equilibrium  Gross-Pitaevskii equation  for the condensate wave function  $\Psi(\bm{r},t)$\cite{Carusotto:13}
  \begin{eqnarray}
\label{eq:gpnl}
 i \hbar \frac{\partial \Psi(\bm{r},t) }{\partial
t} =  & - & \frac{\hbar^{2}}{2 m} \Delta \Psi(\bm{r},t)  +
U(\bm{r},t)\Psi(\bm{r},t) +   g \Psi(\bm{r},t) |\Psi (\bm{r},t)|^{2}
\nonumber \\
&  - &  i{\hbar \over 2 \tau}  \Psi (\bm{r},t)  + i P(\bm{r}),
\end{eqnarray}
where  $m$ is the polariton mass, $\bm{r}=(x,y)$ is a
two-dimensional vector in the plane of the microcavity, and time
$t$, $g$ is  the polariton-polariton interaction strength,  $\tau$
is the polariton lifetime, and  the source terms $P(\bm{r})$
describes incoherent  laser pumping of the polariton reservoir. In
our simulations, we set $\tau = 100$ ps as a representative
value.\cite{Amo:09a,Nelsen:13}

The effective potential for the polaritons
\begin{equation}
U(\bm{r},t) = U_{conf}(\bm{r}) + U_w(\bm{r}) + U_{drag}(\bm{r},t)
\end{equation}
is the sum of the confining potential owing to microcavity
patterning $U_{conf}(\bm{r})$, a linear potential corresponding to a
constant accelerating force in a wedge-shaped microcavity
$U_w(\bm{r})$, and a time-dependent drag $U_{drag}(\bm{r},t)$ caused
by the driving electric current. The confining potential
$U_{conf}(\bm{r})$  that forms the Y-shaped channel is shown in
Fig.~\ref{fig1}b. The potential depth compared to zero value outside
the waveguide is taken equal 1 eV that is a representative value for
waveguides in the
microcavities.\cite{Utsunomiya:08,Daif:06,Kaitouni:06} The average
force acting upon a polariton wave packet in a wedge-shaped
microcavity is  $\bm{F}(\bm{r}) = - \nabla E_{\rm C}(\bm{r})$, where
$E_{\rm C}(\bm{r})$ is the energy of the polariton band taken at the
in-plane  wavevector of the polariton $\bm{k}=0$. For the wedge-like
microcavity considered  in this paper, the energy $E_C(\bm{r})$ is a
linear function of the spatial coordinate\cite{Sermage:01,Nelsen:13}
thus, the force is coordinate-independent. The corresponding
potential is
\begin{equation}
U_{w}(\bm{r}) = - F_x x,
\end{equation}
where $F_x = |\partial E_{\rm C}(\bm{r})/\partial x|$, and we
suppose that the force is applied in $x$-direction, along the stem
of the Y-shaped channel.

The potential that described the drag force in $y$-direction is
taken equal
\begin{equation}
U_{drag}(\bm{r},t) = F_{drag}(t) y
\end{equation}
 in a stripe 400 $\mu$m $<x<600$ $\mu$m
and $U_{drag}(\bm{r},t) = 0$ otherwise. The drag force exerted on
polaritons by charges moving in a neighboring quantum well is
estimated in the $\tau$-approximation as $F_{drag}(t) =\langle p
\rangle / \tau_p = m \gamma {E}(t) / n_n \tau_p$, where $\langle p
\rangle$ is the average gain of the linear momentum of polaritons
owing to the drag, $\tau_p$ is the polariton momentum relaxation
time, ${E}(t)$ is a time-dependent electric field applied in the
plane of the quantum well with free electrons,
 $n_n = 3 \zeta(3) s (k_B T)^3 / 2 \pi \hbar^2 c_s^4 m$ is the density of the normal component in a polariton superfluid,\cite{Berman:12a}
$\zeta(3) \approx 1.202$ is the Riemann zeta function, $s=4$ is the
spin degeneracy factor, $k_B$ is the Boltzmann constant, $T$ is
temperature, and $c_s=\sqrt{g n / m}$ is the sound velocity  in the
polaritonic system. By taking the polariton condensate density
$n=10^{14}$ m$^{-2}$, the separation between the graphene layer with
the excitons and the quantum well with the electrons $D = 15 \
\mathrm{nm}$, $\gamma=6\times 10^{16}$ (Vs)$^{-1}$ and the
relaxation time $\tau_p=6\times 10^{-11}$~s
 as representative parameters for  temperature $T=20$ K,\cite{Zinovev:83,Takagahara:03,Basu:91,Berman:10b}
 one obtains the maximum drag force $F_{drag}=0.3-6$ meV/mm for the working range of electric fields ${E}=10-2\times 10^2$ mV/mm.
 In the simulations, we varied the maximum drag force $F_{drag}$ from 1 to 4 meV/mm.

Formation of the excitons in graphene requires a gap in the electron
and hole excitation spectra, which can be created and dynamically
tuned   by applying an external electric field in the direction
normal to the graphene layer.\cite{Kuzmenko:09, Mak:09, Zhang:09}
The gap can also be opened by chemical doping of
graphene.\cite{Haberer:10} The polariton mass $m$ and the
interaction strength $g$ in gapped graphene depend on the gap energy
$\delta$. The polariton effective mass in gapped graphene
is\cite{Berman:12c}
\begin{equation}
m = 2 (m_{\rm ex}^{-1} + c L_C / \sqrt{\varepsilon} \pi \hbar)
^{-1}, \label{eq:graphm}
\end{equation}
where $ m_{\rm ex}$ is the exciton effective mass in graphene, $L_C$
is the length of the microcavity, $\varepsilon$ is the dielectric
constant of the microcavity, and $c$ is the speed of light in
vacuum. In the simulations, we set $\varepsilon=13$ for a GaAs-based
microcavity.\cite{Berman:10b} We  consider the case of zero detuning
where the cavity photons and the excitons in graphene are in
resonance at $\bm{k}=0$. In this case the length of the microcavity
is\cite{Berman:12c}
\begin{equation}
 L_C = \frac{\hbar \pi c}{(2 \delta - V_0 + C / \delta^2)\sqrt{\varepsilon}}, \label{eq:lc}
\end{equation}
where $V_0 = e^2 / 4 \pi \varepsilon \varepsilon_0 r^{\prime}$, $C =
(\hbar e v_F)^2 / 8 \pi \varepsilon \varepsilon_0 r^{\prime 3}$,
$\varepsilon_0$ is the permittivity of free space, and $v_F\approx
10^6$ m/s is the Fermi-velocity of the electrons in graphene. The
parameter $r^{\prime}$ is found from the  equation
\begin{equation}
2 \delta^2(2 \delta - \hbar \omega) r^{\prime 3} - 2 D \delta^2
r^{\prime 2} + D(\hbar v_F)^2 = 0,
\end{equation}
where $D = e^2 / 4 \pi \varepsilon \varepsilon_0$. For dipolar
excitons in GaAs/AlGaAs coupled quantum wells, the energy of the
recombination peak is $\hbar \omega = 1.61$ eV \cite{Negoita:99}. We
expect similar photon energies in graphene. However, its exact value
depends on the graphene dielectric environment and substrate
properties. The exciton effective mass in Eq. (\ref{eq:graphm})
is\cite{Berman:12c}
\begin{equation}
  m_{\rm ex} = \frac{2 \delta^4}{Cv_F^2}. \label{eq:mex}
\end{equation}
The polariton-polariton interaction strength is
\cite{Berman:12c,Ciuti:03}
\begin{equation}
g = {3 e^2 a_B \over 8 \pi \varepsilon_0 \varepsilon},
\label{eq:graphg}
\end{equation}
where $a_B = 2 \pi \varepsilon_0 \varepsilon  \hbar^2 / m_r e^2$ is
the two-dimensional Bohr radius of the exciton, $m_r = \frac{1 }{ 4}
m_{\rm ex}$ is the exciton reduced mass.

\section{Results and discussion}\label{sec:theory}

\subsection{Electrically controlled optical switch in a wedge-shaped microcavity}

We study propagation of the polariton condensate in a patterned
microcavity by numerically integrating the non-equilibrium
Gross-Pitaevskii equation (\ref{eq:gpnl}) for the wave function
$\Psi(\bm{r},t)$ of the polariton condensate in the microcavity.
 In the model that is described in details in Sec.~\ref{sgpe} we
 we determine the conditions that enable one to control
the polariton spreading in a potential landscape.

The layered semiconductor structure considered in this section is
shown in Fig.~\ref{fig1}a. The excitons
 are positioned in the graphene layer  whereas the free electrons are located in the neighboring, parallel quantum well.
The heterostructure is placed between two high-quality multilayer
mirrors (distributed Bragg reflectors, or DBR). In what follows, we
consider a microcavity decorated with a Y-shaped pattern.
Fig.~\ref{fig1}b demonstrates the potential-energy landscape for the
polaritons due to patterning of the microcavity. Additionally, the
reflectors form a wedge opened towards the positive direction of the
$x$-axis (that is, to the right in Fig.~\ref{fig1})
 that results in a constant force exerted on the polaritons in a microcavity, as explained in
 Sec.~\ref{sgpe}.
In the simulation, we set the constant force equal $F_x=13$ meV/mm
that corresponds
 to a representative experimental value.\cite{Sermage:01,Nelsen:13}
The polaritons are created by an external laser radiation in the region of the stem.
In what follows, an excitation spot has a Gaussian profile with a full width at half maximum  (FWHM) of 65 $\mu$m,
centered in the stem part of the channel at
$\bm{r}=(250,250)$~$\mu$m.

  \begin{figure*}[t]
\includegraphics[width=7cm]{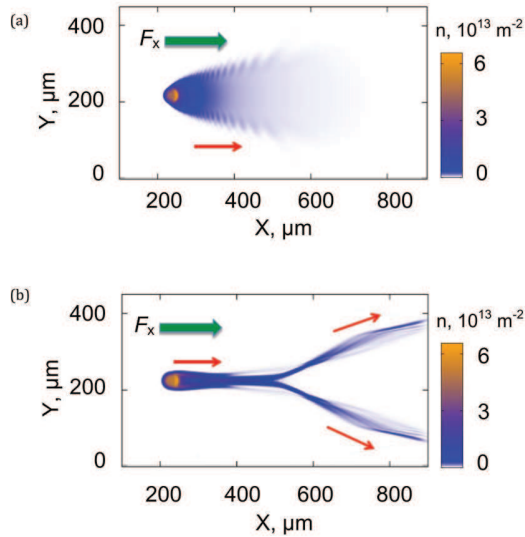}
\caption{Propagation of the polariton condensate in a wedge-shaped
microcavity with embedded graphene (a) with  and (b) without the
pattern deposited on the Bragg reflectors.
 The constant force acting on the polaritons in the $x$-direction owing to the wedge is
$F_x=13$ meV/mm. The color bar shows the polariton density
$n=|\Psi(\bm{r},t)|^2$. The figure shows the steady state reached
after the polariton source is turned on. The polariton flux splits
between the upper and lower branches of the channel in Figure (b).
Red arrows show the  direction of the polariton flow.} \label{fig2}
\end{figure*}
\begin{figure*}[t]
\includegraphics[width=7cm]{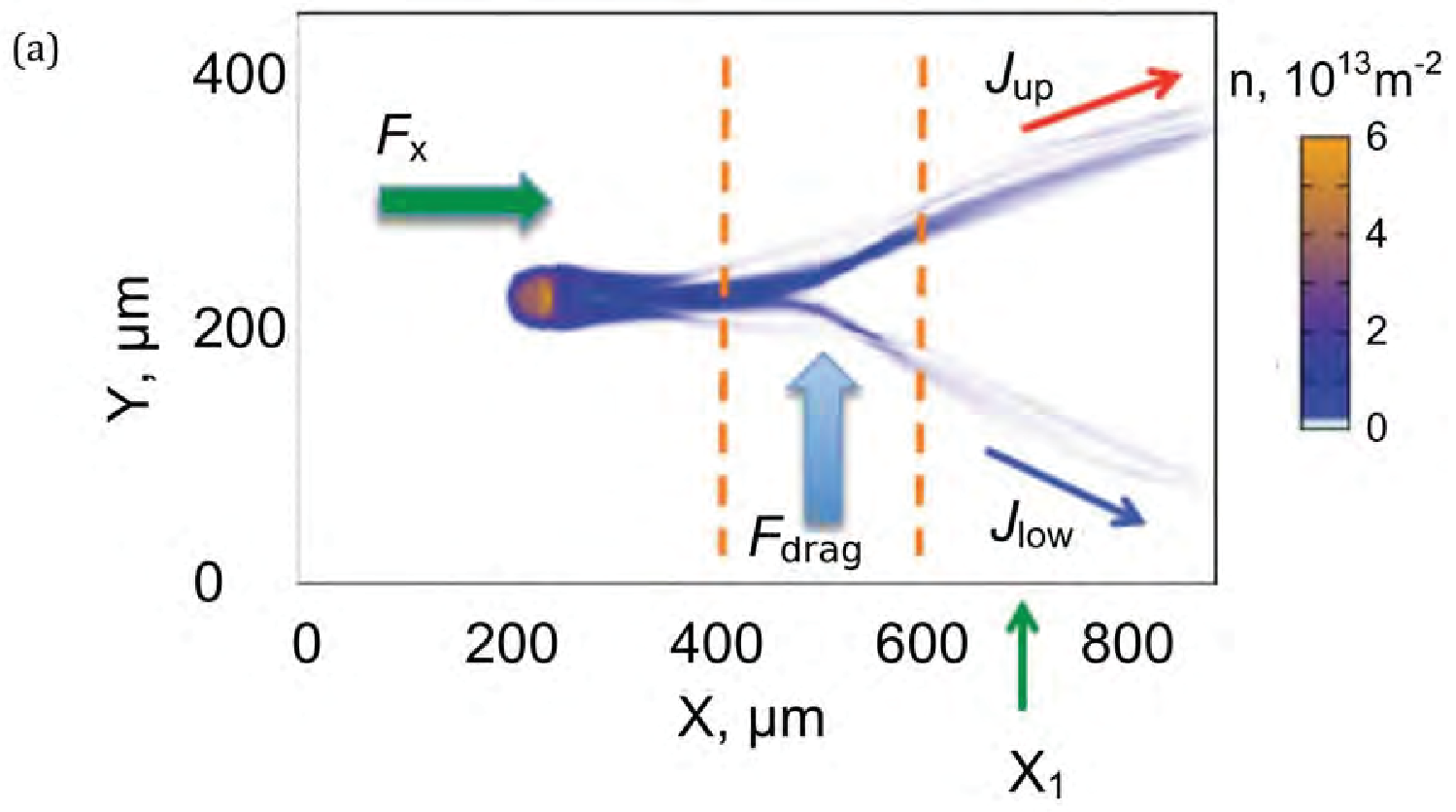}\\
\hspace{-1.5cm}
\includegraphics[width=7cm]{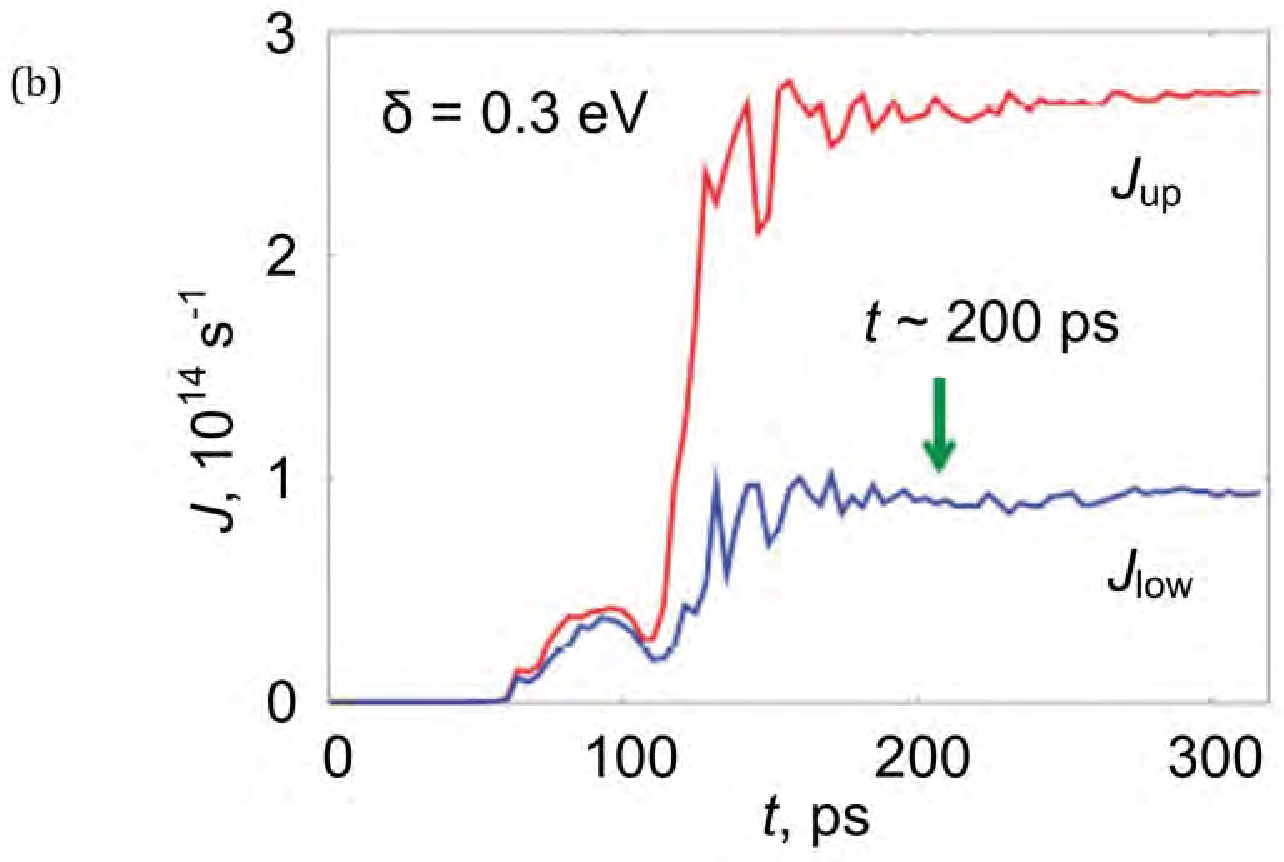}
\caption{(a) Redistribution of the polariton flux between the upper
and lower branches of the channel in response to the driving drag
force $F_{drag}$ in the region of the junction. The boundaries of a
stripe 400 $\mu$m $< x < 600$ $\mu$m in which the drag force is
exerted on the polaritons are shown by vertical dashed lines. The
cross-section of the branches, at which the polariton flux $J$ in
Eq.~(\ref{eq:j}) is determined, is labeled as $x_1$. (b) Dependence
of the polariton flux in the upper $J_{up}$ and lower $J_{low}$
branches on time after the polariton source is turned on. The
transient oscillations last for $t\sim 200$ ps after the source is
switched on and then, the system comes to a steady state. The energy
gap in the graphene layer in the microcavity is $\delta=0.3$ eV. }
\label{fig3}
\end{figure*}

 The effect of  patterning on the polariton condensate propagation is shown in Fig.~\ref{fig2}.
It is seen  that without the pattern (that is, without the confining
potential in the $(x,y)$ plane), a polariton condensate driven by
the constant force $F_x$ forms a wide trail in the direction of the
force with the length of $\sim 400$~$\mu$m. At larger distances from
the excitation spot the condensate density rapidly decreases due to
finite lifetime of polaritons.   Propagation of the condensate under
the same conditions in the presence of a Y-shaped pattern is shown
in Fig.~\ref{fig2}b. As is seen in this Figure, the polariton flow
propagating in the patterned microcavity is localized in the channel
formed by the pattern. In the region of the junction $x \approx
500$~$\mu$m, the polariton condensate flow splits between the
``upper'' and ``lower'' branches of the Y-channel. The polariton
density in the channel at large distances $x\sim 500$ $\mu$m from
the source is significantly higher than that in the microcavity
without the channel (cf.\ Fig.~\ref{fig2}a and b). The increase of
the polariton density is caused by the confining potential that
prevents the polariton flow from spreading in the   $(x,y)$ plane.

To probe the effect of the drag force on the polariton flow
redistribution between the branches, in addition to the constant
force $F_x$ we applied the force $F_{drag}$ directed in
$y$-direction in a stripe  400 $\mu$m $<x<600$ $\mu$m in the region
of the junction. In Fig.~\ref{fig3}a the area, in which the drag
force $F_{drag}$ is exerted, is bounded by the dashed lines. We
found in the simulations that, due to the drag, the polaritons tend
to propagate in the channel in the direction of the drag force
$F_{drag}$ and therefore, they  mostly move in the upper branch of
the channel in Fig~\ref{fig3}a. To characterize the redistribution
of the polaritons between the upper and lower branches we calculated
the total polariton flux  propagating in the branch,
\begin{equation}
J=\int ds j_{\nu}(\bm{r}, t), \label{eq:j}
\end{equation}
where
$j_{\nu}(\bm{r}, t) \equiv \bm{j}(\bm{r}, t) \cdot \bm{\nu}$ is the component of the polariton flux density
parallel to the channel, $\bm{\nu}$ is a unit vector along the branch,  and
\begin{equation}
\bm{j}(\bm{r}, t) = {\hbar \over 2 i m} (\Psi^*(\bm{r},t) \nabla \Psi(\bm{r},t) - \Psi(\bm{r},t) \nabla \Psi^*(\bm{r},t))
\end{equation}
is the flux density.\cite{Messiah:61} We integrate  in
Eq.~(\ref{eq:j}) over the cross section of the branch. The flux $J$
was calculated at $x=x_1 \equiv 700$~$\mu$m that is, at a distance
from the junction much larger than the width of the channel
$a_{ch}\sim 60$ $\mu$m.

\begin{figure*}[t]
\includegraphics[width=7cm]{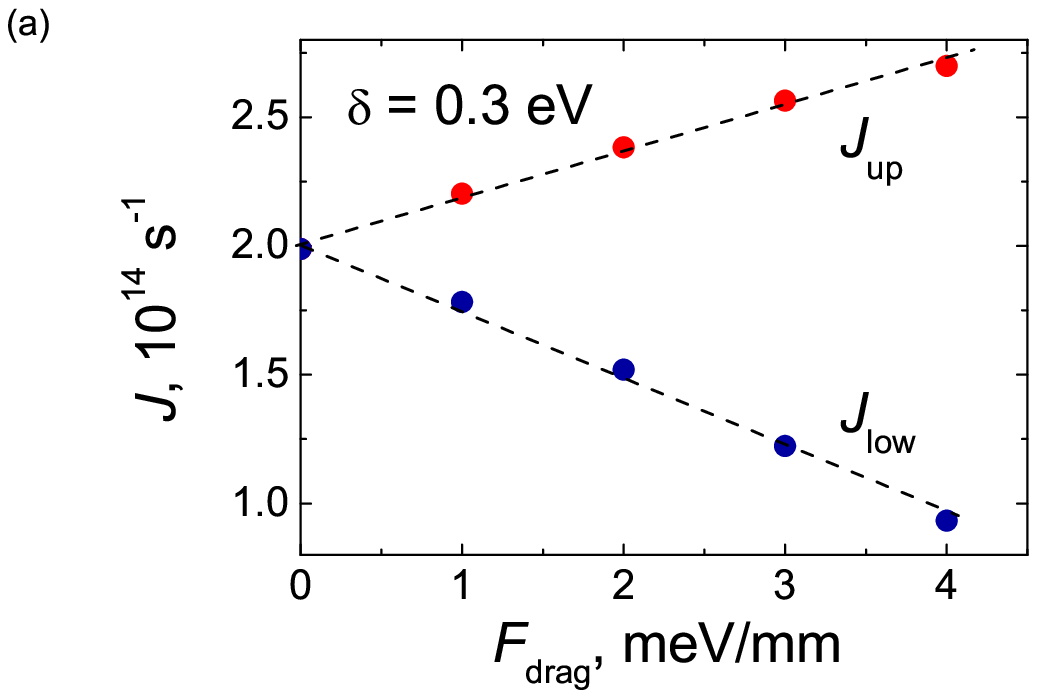}\\
\includegraphics[width=7cm]{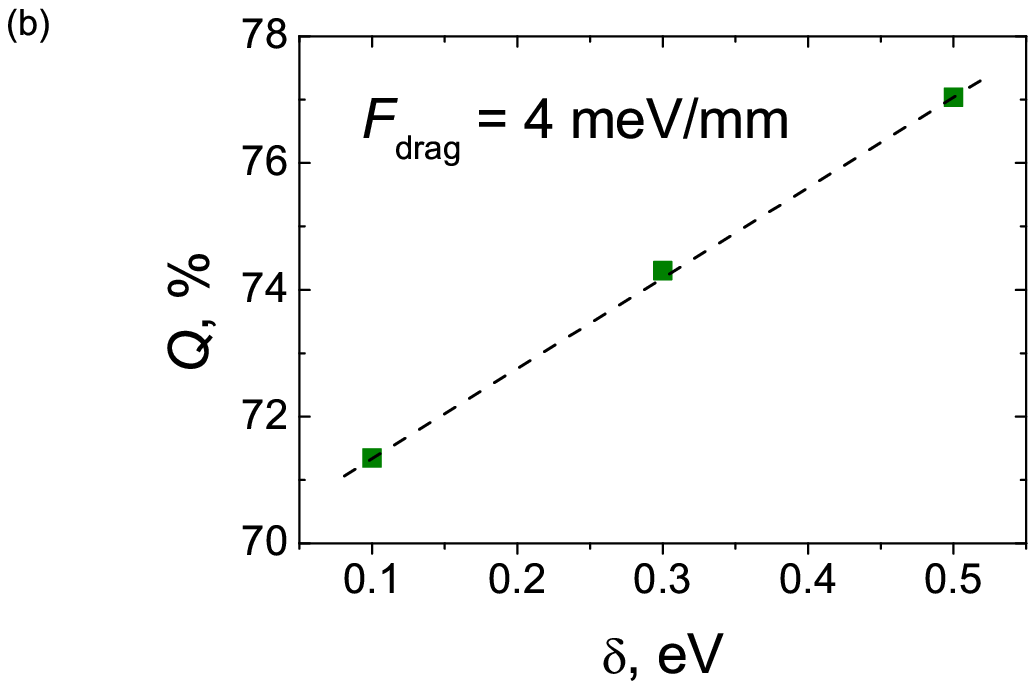}
\caption{(a) Redistribution of the polariton flux between the upper
and lower branches in response to the magnitude of the driving drag
force $F_{drag}$. The polaritons fluxes in the upper and lower
channel, $J_{up}$ and $J_{low}$, are determined at the cross-section
labeled as $x_1$  in Fig.~\ref{fig3}a in a steady state after the
transient oscillations damped. (b) Performance of the system Q,
Eq.~(\ref{eq:q}), as a function of the energy gap in the graphene
layer. It is seen that up to 77\% of polaritons are propagating
through the upper channel if the drag force   $F_{drag}=4$~meV/mm is
applied at the junction. } \label{fig4}
\end{figure*}

 Fig.~\ref{fig3}b shows the dependencies of the total polariton flux
trough the upper (red curve) and lower (blue curve) branches  as functions of time $t$.
 In these simulations, the energy gap in the graphene quasiparticle excitation spectrum is set
$\delta=0.3$ eV, the drag force is $F_{drag}=4$ meV/mm, and the
polariton source is turned on at the moment $t=0$. According to the
results shown in Fig.~\ref{fig3}b,  the polaritons propagate to the
point $x_1$ at $t\sim 70$ ps after the source is turned on. During
the time interval 70 ps $< t< 200$ ps the polariton flux exhibits
transient oscillations and then, tends to a constant value at $t
\geq 200$ ps. As it follows from Fig.~\ref{fig3}b, the flux through
the upper branch of the channel in the steady state is $J\approx 2.7
\times 10^{14}$ s$^{-1}$, that is $\approx 2.8\times$ larger than
that through the lower branch. Therefore, the polariton flow is
redistributed among the branches  in response to the drag force
owing to the external, driving electric current. To  quantify the
redistribution of the polariton flux in the channel, we studied the
dependence of the flux $J$ on the drag force $F_{drag}$. In the
simulations, we varied $F_{drag}$ from 1 meV/mm to 4 meV/mm, as
explained in Sec.~\ref{sgpe}. It is shown in Fig.~\ref{fig4}a that
the flux in the upper branch $J_{up}$ gradually increases with the
rise of the drag force $F_{drag}$. In its turn, the flux through the
lower channel, $J_{low}$ decreases thus, the total flux $J_{tot}=
J_{up} + J_{low}$ remains approximately constant. To more fully
characterize the  polariton flow in the channel,  we determined the
performance of the system,
\begin{equation}
Q = (J_{up} / J_{tot}) \times 100\%, \label{eq:q}
\end{equation}
 as a function
of the energy gap in the graphene layer.  We varied the gap $\delta$
from   0.1 to 0.5 eV that corresponds to the experimentally
accessible range.\cite{Kuzmenko:09, Mak:09, Zhang:09,Haberer:10} The
results for the constant drag force $F_{drag}=4$ meV/mm are
summarized in Fig.~\ref{fig4}b. It is seen that  the performance
 $Q$ can be approximated by a linear function of the gap $\delta$ and it reaches $\sim 77$\% for the maximum gap $\delta=0.5$ eV.
In other words, more than $3/4$ of the total polariton flux propagates through the upper channel for $F_{drag}=4$ meV/mm, and less than $1/4$
of the flux only propagates through the lower channel.

Summarizing, we demonstrated that one can govern the propagation of
a polariton condensate  in  the Y-shaped channel by means of the external electric current running
in a neighboring quantum well.
This make it possible to construct an optical, polariton-based switch controlled by an external voltage.

\begin{figure*}[t]
\includegraphics[width=7cm]{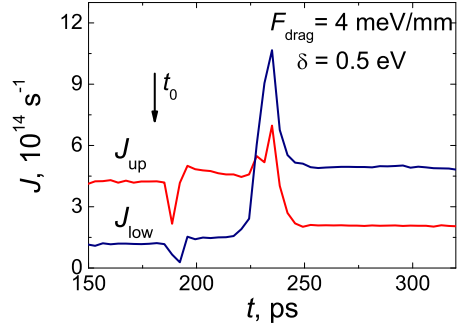}
 \caption{Response of the system to switching  direction of the
drag force. The drag is directed towards the upper channel at $t<t_0
\equiv 180$ ps after the source is turned on; the polaritons mostly
propagate via the upper branch. The direction of the drag is changed
to opposite at $t=t_0$ and then, the transient oscillations of the
polariton flux  occur for $\sim 70$ ps. After the oscillations are
damped, most of polaritons propagate via the lower branch.  The
magnitude of the drag force and the energy gap are labeled in the
figure.} \label{fig5}
\end{figure*}

 Finally, to estimate the response speed  of the system,
 we studied the transient dynamics of a polariton condensate in the channel when the direction of the driving current
 is inverted. In these simulations,  we turned on the pumping and waited $t\sim 180$ ps
 until the system reached an equilibrium in the presence of a drag force.
 In this steady state, the polaritons  mostly propagated  in the upper channel, in accordance with
 our previous consideration.
 At $t=180$ ps, we changed the direction  of the drag force to  opposite.
 Since the drag force now pushes the polaritons down in the junction area, the polaritons
 tend to propagate via the lower channel, whereas the flux through the upper channel now decreases.
 After the transient oscillations were damped, the system came to a new equilibrium state where most of the polaritons propagated
 through the lower channel. The dependencies of the polariton fluxes $J_{up}$ and $J_{low}$ on time are shown
 in Fig.~\ref{fig5}. According to this Figure,  the characteristic switching time of the system between the states where the polaritons
 propagate via the upper and lower branches is $t_{sw}\approx 70$ ps. This means
 that the maximum switch frequency for this junction  can reach $f = t_{sw}^{-1} \approx 14$ GHz
 at given linear dimensions of the system.
 This frequency, however,   should rise with decreasing channel length.
 In this paper we only consider the switch of length
 $L\sim 600$ $\mu$m and defer the size-dependent effects to the future studies.



\subsection{Polariton switch in a flat microcavity}

We now consider another design of the  polariton switch based on
the drag effect. In this case, the reflectors are parallel to each other.
In such ``flat'' microcavity  the only force acting on the polaritons is the drag force
due to the interactions with the current.
Such flat microcavities are widely used in the studies of exciton polaritons
in semiconductor heterostructures.\cite{Balili:09,Butov:03,Lai:07,Assmann:11,Kasprzak:06,Wu:14,Furchi:12,Youngblood:14}

In this design, the microcavity has no decoration. The Y-shaped
channel is formed by graphene stripes.\cite{Berger:06} While the
design of nanojunction of the graphene stripes is technologically
challenging, the possibility of  fabrication of a junction of
graphene stripes with help of the covalent functionalization has
recently been proposed.\cite{Cocchi:11} We note that the utility of
electric splitters  based on graphene structures has recently been
discussed in the literature.\cite{Zhu:13} The current-carrying
quantum well is now Y-shaped that is, it repeats the shape of the
channel. Similarly to the design above, the excitons in graphene are
excited due to the laser pumping at the stem of the channel. The
main concept is schematically illustrated in Fig.~\ref{fig6}.

The voltage $V$ is applied across the electron QW between  the stem
and one of the branches. The polaritons formed by excitons in
graphene and cavity photons are dragged by the electron current in
the electron QW along one of the branches. The direction of the
electric current and, hence, of the polariton flux is defined by the
electric switch $S$ (see Fig.~\ref{fig6}). If  the switch $S$ is
closed in position A then,  the electric current in the electron QW
runs in the upper branch. Due to the drag effect caused by the
interactions between the excitons in graphene and the electrons in
QW,\cite{Berman:10a,Berman:10b} the electron current induces the
flow of polariton quasiparticles in the cavity along  the upper
branch. Then, if the switch is closed in position $B$,
 the flow of polariton quasiparticles  propagates along the lower branch.
Thus, the path of the polaritons in the Y-shaped channel can be dynamically changed
by changing the position of the switch $S$.

The flux density $\bm{j}$ of the polariton superfluid in the channel
is given by Eq.~(\ref{eq:j0}). For the polaritons velocities smaller
than the speed of sound $c_s$ in the polariton superfluid, the
polariton flux density can be found as follows\cite{Berman:10b}
\begin{eqnarray}
\bm{j}\equiv n_{n} \bm{v} = \gamma \bm{E} \ ,
\end{eqnarray}
where $n_{n}$ is the density of the normal component in the
polariton subsystem formed by the quasiparticles, and $\bm{v}$
is the average velocity of the dragged polariton quasiparticles. Estimating the magnitude of the electric field as
$|\bm{E}| \simeq
V/L$, where $V$ is the voltage across the channel and $L$ is the
length of the channel, one obtains for the magnitude  of the averaged polariton velocity
\begin{eqnarray}
\label{v} v = \frac{\gamma V}{n_{n}L} \ .
\end{eqnarray}
The time $t$, required for a quasiparticle to
propagate across the channel, is estimated as
\begin{eqnarray}
\label{t} t \simeq \frac{L}{v} = \frac{n_{n}L^{2}}{\gamma V}.
\end{eqnarray}

In the relevant parameter range discussed in Sec.~\ref{sgpe},
Eq.~(\ref{t}) gives the estimate $t\approx72$ ps for the channel of
length $L=100$ $\mu$m,
 the voltage $V=20$ mV, and energy gap $\delta=0.4$ eV.
The average velocity of the polaritons  calculated from
Eq.~(\ref{v}) is $v= 1.38 \times 10^{6}$ m/s that is smaller than
the sound velocity  $c_{s} = 1.81 \times 10^{6}$ m/s in the
polariton superfluid. Therefore, the breakdown of  superfluidity in
the system does not occur.\cite{Berman:10b,Wouters:10a} For
polaritons with lifetime  $\tau \sim100$ ps,\cite{Amo:09a,Nelsen:13}
the time $t$ required for the quasiparticles to propagate across the
channel is smaller than $\tau$  that validates the proposed design.

\section{Conclusions}
In the above studies we considered two designs of an electrically-controlled optical switch.
In both cases, we focused on propagation of  polaritons in a microcavity, which
are a quantum superposition of cavity photons and excitons.
A significant challenge is
designing a system where the direction of the polariton flow is controlled by means of the electrical current since
both photons and excitons  are electrically neutral.
In our studies, to control the polariton propagation we utilize the drag effect that is, the entrainment
of polaritons by an electric current running in the neighboring quantum well.
The drag effect for excitons in a planar geometry is well studied. Recently, the drag of
polaritons by the electric current in an unrestricted planar microcavity has been considered theoretically.
In the above studies, we utilize this approach to design nanostructures, in
which the  polaritons are localized in
a circuit made by quasi-one dimensional polariton wires.

In the first setup, the polaritons propagate in the Y-shaped channel  owing to a constant force created in a wedge-shaped microcavity.
In this case, the controlling drag force is directed perpendicularly to the direction of the stem of the channel and it
pushes the polariton in one of the two branches. Changes in the direction of the electric current results in switching of the path
in which the polaritons propagate.

In the second setup, the polaritons in a flat microcavity are considered. In this case, the constant force is absent and the polaritons
  are  only dragged by an electric current running in a Y-shaped quantum well.
The channel containing the excitons is formed by  graphene stripes embedded in the microcavity.
Despite the latter setup is more technologically challenging, mostly due to the requirement of a high-quality junction
between the stripes, this scheme provides additional advantages for the polariton flow control.
Specifically, the magnitude of the polariton flow  can be tuned by setting the voltage across the channel to a given value.

In both cases, we focused on microcavities with embedded gapped graphene.
The advantage of graphene, compared to standard semiconductor quantum wells, is in the possibility to dynamically
tune the effective mass of the polaritons by application of an electric field normal to the microcavity plane.
This provides  an additional opportunity to control the polariton propagation in the device.



\acknowledgments

The authors are gratefully acknowledge support from Army Research Office, grant \#64775-PH-REP.
G.V.K. is also grateful for support to Professional Staff Congress -- City University of New York, award  \#66140-0044.
The authors are  grateful to the Center for Theoretical Physics
of New York City College of Technology of the City University of New York
for providing computational resources.


\end{document}